\documentclass[%
 reprint,
superscriptaddress,
showpacs,preprintnumbers,
 aps,
]{revtex4-1}

\usepackage{graphicx}
\usepackage{dcolumn}
\usepackage{bm}
\usepackage{amssymb}

\begin{document}

\preprint{APS/123-QED}

\title{Insight into the narrow structure in {\boldmath{$\eta$}}-photoproduction on the neutron
from helicity dependent cross sections}


\author{L.~Witthauer}
\affiliation{Department of Physics, University of Basel, Basel, Switzerland}
\author {M.~Dieterle}
\affiliation{Department of Physics, University of Basel, Basel, Switzerland}
\author{S.~Abt}
\affiliation{Department of Physics, University of Basel, Basel, Switzerland}
\author{P.~Achenbach}
\affiliation{Institut f\"ur Kernphysik, University of Mainz, Mainz, Germany}
\author{F.~Afzal}
\affiliation{Helmholtz-Institut f\"ur Strahlen- und Kernphysik, University of Bonn, Bonn, Germany}
\author{Z.~Ahmed}
\affiliation{University of Regina, Regina, SK S4S 0A2 Canada}
\author{J.R.M.~Annand}
\affiliation{SUPA School of Physics and Astronomy, University of Glasgow, Glasgow, G12 8QQ, UK}
\author{H.J.~Arends}
\affiliation{Institut f\"ur Kernphysik, University of Mainz, Mainz, Germany}
\author{M.~Bashkanov}
\affiliation{SUPA School of Physics, University of Edinburgh, Edinburgh EEH9 3JZ, UK}
\author{R.~Beck}
\affiliation{Helmholtz-Institut f\"ur Strahlen- und Kernphysik, University of Bonn, Bonn, Germany}
\author{M.~Biroth}
\affiliation{Institut f\"ur Kernphysik, University of Mainz, Mainz, Germany}
\author{N.S.~Borisov}
\affiliation{Joint Institute for Nuclear Research,141980 Dubna, Russia}
\author{A.~Braghieri}
\affiliation{INFN Sezione di Pavia, Pavia, Italy}
\author{W.J.~Briscoe}
\affiliation{Center for Nuclear Studies, The George Washington University, Washington, DC, USA}
\author{ F.~Cividini}
\affiliation{Institut f\"ur Kernphysik, University of Mainz, Mainz, Germany}
\author{S.~Costanza}\altaffiliation{Also at Dipartimento di Fisica, Universit\`a di Pavia, Pavia, Italy.}
\affiliation{INFN Sezione di Pavia, Pavia, Italy}
\author{C.~Collicott}
\affiliation{Department of Astronomy and Physics, Saint Marys University, Halifax, Canada}
\author{A.~Denig}
\affiliation{Institut f\"ur Kernphysik, University of Mainz, Mainz, Germany}
\author{E.J.~Downie}
\affiliation{Institut f\"ur Kernphysik, University of Mainz, Mainz, Germany}
\affiliation{Center for Nuclear Studies, The George Washington University, Washington, DC, USA}
\author{P.~Drexler}
\affiliation{Institut f\"ur Kernphysik, University of Mainz, Mainz, Germany}
\author{M.I.~Ferretti-Bondy}
\affiliation{Institut f\"ur Kernphysik, University of Mainz, Mainz, Germany}
\author{S.~Gardner}
\affiliation{SUPA School of Physics and Astronomy, University of Glasgow, Glasgow, G12 8QQ, UK}
\author{S.~Garni}
\affiliation{Department of Physics, University of Basel, Basel, Switzerland}
\author{D.I.~Glazier}
\affiliation{SUPA School of Physics and Astronomy, University of Glasgow, Glasgow, G12 8QQ, UK}
\affiliation{SUPA School of Physics, University of Edinburgh, Edinburgh EEH9 3JZ, UK}
\author{D.~Glowa}
\affiliation{SUPA School of Physics, University of Edinburgh, Edinburgh EEH9 3JZ, UK}
\author{W.~Gradl}
\affiliation{Institut f\"ur Kernphysik, University of Mainz, Mainz, Germany}
\author{M.~G\"unther}
\affiliation{Department of Physics, University of Basel, Basel, Switzerland}
\author{G.M.~Gurevich}
\affiliation{Institute for Nuclear Research, Moscow, Russia}
\author{D.~Hamilton}
\affiliation{SUPA School of Physics and Astronomy, University of Glasgow, Glasgow, G12 8QQ, UK}
\author{D.~Hornidge}
\affiliation{Mount Allison University, Sackville, New Brunswick E4L 1E6, Canada}
\author{G.M.~Huber}
\affiliation{University of Regina, Regina, SK S4S 0A2 Canada}
\author{A.~K{\"a}ser}
\affiliation{Department of Physics, University of Basel, Basel, Switzerland}
\author{V.L.~Kashevarov}
\affiliation{Institut f\"ur Kernphysik, University of Mainz, Mainz, Germany}
\author{ S.~Kay}
\affiliation{SUPA School of Physics, University of Edinburgh, Edinburgh EEH9 3JZ, UK}
\author{I.~Keshelashvili}\altaffiliation{Now at Institut f\"ur Kernphysik, FZ J\"ulich, 52425 J\"ulich, Germany}
\affiliation{Department of Physics, University of Basel, Basel, Switzerland}
\author{R.~Kondratiev}
\affiliation{Institute for Nuclear Research, Moscow, Russia}
\author{M.~Korolija}
\affiliation{Rudjer Boskovic Institute, Zagreb, Croatia}
\author{B.~Krusche}\email[]{Corresponding author: email bernd.krusche@unibas.ch}
\affiliation{Department of Physics, University of Basel, Basel, Switzerland}
\author{A.B.~Lazarev}
\affiliation{Joint Institute for Nuclear Research,141980 Dubna, Russia}
\author{J.M.~Linturi}
\affiliation{Institut f\"ur Kernphysik, University of Mainz, Mainz, Germany}
\author{V.~Lisin}
\affiliation{Institute for Nuclear Research, Moscow, Russia}
\author{K.~Livingston}
\affiliation{SUPA School of Physics and Astronomy, University of Glasgow, Glasgow, G12 8QQ, UK}
\author{S.~Lutterer}
\affiliation{Department of Physics, University of Basel, Basel, Switzerland}
\author{I.J.D.~MacGregor}
\affiliation{SUPA School of Physics and Astronomy, University of Glasgow, Glasgow, G12 8QQ, UK}
\author{J.~Mancell}
\affiliation{SUPA School of Physics and Astronomy, University of Glasgow, Glasgow, G12 8QQ, UK}
\author{D.M.~Manley}
\affiliation{Kent State University, Kent, OH, USA}
\author{P.P.~Martel}
\affiliation{Institut f\"ur Kernphysik, University of Mainz, Mainz, Germany}
\author{V.~Metag}
\affiliation{II. Physikalisches Institut, University of Giessen, Germany}
\author{W.~Meyer}
\affiliation{Institut f\"ur Experimentalphysik, Ruhr Universit\"at, 44780 Bochum, Germany}
\author{R.~Miskimen}
\affiliation{University of Massachusetts Amherst, Amherst, Massachusetts 01003, USA}
\author{E.~Mornacchi}
\affiliation{Institut f\"ur Kernphysik, University of Mainz, Mainz, Germany}
\author{A.~Mushkarenkov}
\affiliation{Institute for Nuclear Research, Moscow, Russia}
\affiliation{University of Massachusetts Amherst, Amherst, Massachusetts 01003, USA}
\author{A.B.~Neganov}
\affiliation{Joint Institute for Nuclear Research,141980 Dubna, Russia}
\author{A.~Neiser}
\affiliation{Institut f\"ur Kernphysik, University of Mainz, Mainz, Germany}
\author{M.~Oberle}
\affiliation{Department of Physics, University of Basel, Basel, Switzerland}
\author{M.~Ostrick} 
\affiliation{Institut f\"ur Kernphysik, University of Mainz, Mainz, Germany}
\author{P.B.~Otte}
\affiliation{Institut f\"ur Kernphysik, University of Mainz, Mainz, Germany}
\author{ D.~Paudyal}
\affiliation{University of Regina, Regina, SK S4S 0A2 Canada}
\author{P.~Pedroni}
\affiliation{INFN Sezione di Pavia, Pavia, Italy}
\author{A.~Polonski}
\affiliation{Institute for Nuclear Research, Moscow, Russia}
\author{S.N.~Prakhov}
\affiliation{Institut f\"ur Kernphysik, University of Mainz, Mainz, Germany}
\affiliation{University of California at Los Angeles, Los Angeles, CA, USA}
\author{A.~Rajabi}
\affiliation{University of Massachusetts Amherst, Amherst, Massachusetts 01003, USA}
\author{G.~Reicherz}
\affiliation{Institut f\"ur Experimentalphysik, Ruhr Universit\"at, 44780 Bochum, Germany}
\author{G.~Ron}
\affiliation{Racah Institute of Physics, Hebrew University of Jerusalem, Jerusalem 91904, Israel}
\author{T.~Rostomyan}\altaffiliation{Now at Department of Physics and Astronomy., Rutgers University,
Piscataway, New Jersey, 08854-8019}
\affiliation{Department of Physics, University of Basel, Basel, Switzerland}
\author{A.~Sarty}
\affiliation{Department of Astronomy and Physics, Saint Marys University, Halifax, Canada}
\author{C.~Sfienti}
\affiliation{Institut f\"ur Kernphysik, University of Mainz, Mainz, Germany}
\author{M.H.~Sikora}
\affiliation{SUPA School of Physics, University of Edinburgh, Edinburgh EEH9 3JZ, UK}
\author{V.~Sokhoyan}
\affiliation{Institut f\"ur Kernphysik, University of Mainz, Mainz, Germany}
\affiliation{Center for Nuclear Studies, The George Washington University, Washington, DC, USA}
\author{K.~Spieker}
\affiliation{Helmholtz-Institut f\"ur Strahlen- und Kernphysik, University of Bonn, Bonn, Germany}
\author{O.~Steffen}
\affiliation{Institut f\"ur Kernphysik, University of Mainz, Mainz, Germany}
\author{I.I.~Strakovsky}
\affiliation{Center for Nuclear Studies, The George Washington University, Washington, DC, USA}
\author{Th.~Strub}
\affiliation{Department of Physics, University of Basel, Basel, Switzerland}
\author{I.~Supek}
\affiliation{Rudjer Boskovic Institute, Zagreb, Croatia}
\author{A.~Thiel}
\affiliation{Helmholtz-Institut f\"ur Strahlen- und Kernphysik, University of Bonn, Bonn, Germany}
\author{M.~Thiel}
\affiliation{Institut f\"ur Kernphysik, University of Mainz, Mainz, Germany}
\author{A.~Thomas}
\affiliation{Institut f\"ur Kernphysik, University of Mainz, Mainz, Germany}
\author{M.~Unverzagt}
\affiliation{Institut f\"ur Kernphysik, University of Mainz, Mainz, Germany}
\author{Yu.A.~Usov}
\affiliation{Joint Institute for Nuclear Research,141980 Dubna, Russia}
\author{S.~Wagner}
\affiliation{Institut f\"ur Kernphysik, University of Mainz, Mainz, Germany}
\author{N.K.~Walford}
\affiliation{Department of Physics, University of Basel, Basel, Switzerland}
\author{D.P.~Watts}
\affiliation{SUPA School of Physics, University of Edinburgh, Edinburgh EEH9 3JZ, UK}
\author{D.~Werthm\"uller}
\affiliation{Department of Physics, University of Basel, Basel, Switzerland}
\affiliation{SUPA School of Physics and Astronomy, University of Glasgow, Glasgow, G12 8QQ, UK}
\author{J.~Wettig}
\affiliation{Institut f\"ur Kernphysik, University of Mainz, Mainz, Germany}
\author{M.~Wolfes}
\affiliation{Institut f\"ur Kernphysik, University of Mainz, Mainz, Germany}
\author{L.~Zana}
\affiliation{SUPA School of Physics, University of Edinburgh, Edinburgh EEH9 3JZ, UK}
\collaboration{A2 Collaboration at MAMI}
%

\date{\today}

\begin{abstract}
The double polarization observable $E$ and the helicity dependent cross sections $\sigma_{1/2}$ 
and $\sigma_{3/2}$ were measured for $\eta$ photoproduction from quasi-free protons and neutrons. 
The circularly polarized tagged photon beam of the A2 experiment at the Mainz MAMI accelerator was 
used in combination with a longitudinally polarized deuterated butanol target. The almost $4\pi$ 
detector setup of the Crystal Ball and TAPS is ideally suited to detect the recoil nucleons 
and the decay photons from $\eta\rightarrow 2\gamma$ and $\eta\rightarrow 3\pi^0$. The results show
that the narrow structure previously observed in $\eta$ photoproduction from the neutron is only 
apparent in $\sigma_{1/2}$ and hence, most likely related to a spin-1/2 amplitude. Nucleon
resonances that contribute to this partial wave in $\eta$ production are only $N1/2^-$ ($S_{11}$)
and $N1/2^+$ ($P_{11}$). Furthermore, the extracted Legendre coefficients of the angular distributions 
for $\sigma_{1/2}$ are in good agreement with recent reaction model predictions assuming a narrow 
resonance in the $P_{11}$ wave as the origin of this structure.
\end{abstract}

\maketitle

Photoproduction of $\eta$ mesons is important for the investigation of the nucleon excitation spectrum.
Due to its isoscalar nature, the $\eta$ only couples to isospin $I=1/2$ $N^{\star}$ resonances.
In the threshold region, this reaction is completely dominated by the excitation of the $N(1535)1/2^-$ 
resonance \cite{Krusche_03} and at higher incident photon energies, contributions from several other 
excited nucleon states have been identified \cite{Krusche_15}. Currently, a large effort is underway
at modern photon-beam facilities (see \cite{Krusche_15} for a recent summary) to study the
$\gamma p\rightarrow p\eta$ reaction using both single and double polarization
observables. However, during the last few years, photoproduction of $\eta$ mesons off the neutron has
attracted additional interest. The reason is the discovery of an unusually narrow structure in the excitation
function at incident photon energies of 1~GeV (corresponding to an $\eta$-neutron invariant mass of
$W\approx 1.67$~GeV). This structure was first observed by the GRAAL collaboration \cite{Kuznetsov_07} 
and confirmed by the CBELSA/TAPS collaboration \cite{Jaegle_08,Jaegle_11a} in Bonn, and
at LNS in Sendai \cite{Miyahara_07}. Recent high-statistics measurements at the MAMI facility in Mainz 
with deuterium and $^3$He targets \cite{Werthmueller_13,Witthauer_13,Werthmueller_14} have extracted a 
position of the narrow structure of $W$ = (1670$\pm$5)~MeV with a width of only $\Gamma$~=~(30$\pm$15)~MeV.
This structure is not observed in $\eta$ photoproduction off the proton \cite{McNicoll_10}. The cross section of 
$\gamma p\rightarrow p\eta$ shows only a small dip at this energy \cite{McNicoll_10,Krusche_15}. 
However, recently, two narrow structures were observed in the beam asymmetry $\Sigma$ of Compton scattering
of the proton \cite{Kuznetsov_15}. One of these structures appears close to the above discussed peak in 
$\eta$ production off neutrons and the other at $W\approx 1.726$~GeV. Meanwhile, a counterpart of
the latter peak was also unambiguously identified in the cross section of the $\gamma n\rightarrow n\eta$
reaction \cite{Werthmueller_15}. 

The nature of these structures has not yet been established. The prominent peak observed in $\eta$ production
off the neutron at $W\approx 1.67$~GeV has been discussed as a new narrow resonance (with exotic properties) 
\cite{Polyakov_03,Arndt_04,Choi_06,Fix_07,Shrestha_12}. It is currently listed in the Review of Particle
Physics (RPP) \cite{PDG_14} as a tentative $N(1685)$ state with unknown spin/parity. However, other works 
suggest coupled-channel effects of known nucleon resonances \cite{Shklyar_07,Shyam_08}, or contributions 
from intermediate strangeness states \cite{Doering_10} as the underlying cause. A fit \cite{Anisovich_15} 
from the BnGa group to the high statistics MAMI deuteron data \cite{Werthmueller_13,Werthmueller_14} suggests 
an interference in the $J^P = 1/2^{-}$ partial wave between contributions from the well-known $N$(1535) and 
$N$(1650) resonances. Fits of these unpolarized data with the BnGa model including a narrow 
$P_{11}$-like $N(1685)$ resonance were seen as inferior \cite{Anisovich_15}. 

The aim of the present work is to determine the relevant partial wave directly from experimental data.  
For this purpose, the double polarization observable $E$ was measured with a longitudinally polarized target
and a circularly polarized photon beam. It is defined as 
\cite{Barker_75}:
\begin{equation}
E= \frac{\sigma_{1/2}-\sigma_{3/2}}{\sigma_{1/2}+\sigma_{3/2}}\, ,
\label{eq:E}
\end{equation}
where $\sigma_{1/2}$ and $\sigma_{3/2}$ are the helicity dependent cross sections with anti-parallel 
or parallel photon and nucleon spin, respectively. 
Nucleon resonances with spin $J=1/2$ contribute only to $\sigma_{1/2}$, while states with spin $J\geq 3/2$ 
can also couple to $\sigma_{3/2}$. Hence, structures in the $S_{11}$ or $P_{11}$ partial waves appear only 
in $\sigma_{1/2}$, but not in $\sigma_{3/2}$. So far, in $\eta$ production, this observable has only been 
explored for the reaction with free protons \cite{Senderovich_16}, for which it turned out to be very
powerful in restricting parameters of reaction model analyses. 

The experiments were performed at the Mainz MAMI accelerator \cite{Walcher_90}. Circularly polarized 
tagged photons \cite{McGeorge_08} were created via the bremsstrahlung process with longitudinally polarized 
($P_{e}\sim80\%$) electrons. The beam helicity was flipped once per second. The polarization of the 
electron beam was measured daily with Mott scattering (after the linac stage of the accelerator at 
electron energies of 3.65~MeV) and constantly monitored with M$\o$ller scattering of the high energy 
electrons from the bremsstrahlung radiator. The polarization of the photon beam was deduced from 
the energy-dependent polarization transfer factors given by Olsen and Maximon \cite{Olsen_59}.
The deuterated butanol (C$_4$D$_9$OD) target was polarized in the longitudinal direction using Dynamic Nuclear 
Polarization \cite{Bradtke_99}. The target polarization was measured before and after data taking using an 
NMR measurement technique and was interpolated by an exponential function. Due to small inhomogeneities of 
the polarizing magnetic field, the target was not homogeneously polarized across its diameter for the initial
beam times (so that the NMR measurements did not correctly reflect the polarization degree in
the target area interacting with the beam). Therefore, results were renormalized to the final data taking 
period for which this problem was resolved. 

The experimental setup combined the Crystal Ball (CB) \cite{Starostin_01} and TAPS \cite{Gabler_94} 
calorimeters with additional detectors for charged particle identification and covered 98\% of $4\pi$.
Detected and analyzed were the photons from the $\eta$ decays (results from $\eta\rightarrow \gamma\gamma$
and $\eta\rightarrow 3\pi^0\rightarrow 6\gamma$ were consistent and have been averaged) and the recoil nucleons.
The detector was identical to the setup used for the measurements with unpolarized targets which is 
discussed in detail in \cite{Witthauer_13,Werthmueller_14}. Also, all analysis procedures were identical 
to those described in these references. This includes the clean identification of $\eta$ production off 
quasi-free nucleons, the Monte Carlo simulations of the detector response, and the reconstruction of  
final-state kinematics used to remove the effects from nuclear Fermi motion. The latter is essential for 
the investigation of narrow structures.

\begin{figure}[b]
\centerline{\resizebox{0.5\textwidth}{!}{\includegraphics{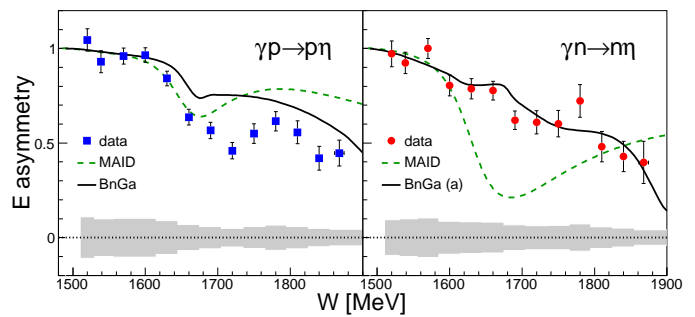}}}
\caption{\label{fig:E} (Color online) Double polarization observable $E$ for $\gamma p\rightarrow p\eta$ 
(left hand side) and $\gamma n\rightarrow n\eta$ (right hand side). 
Gray shaded areas: systematic uncertainties.
Curves: predictions from MAID (green, dashed) \cite{Chiang_02} and BnGa (model based on $S_{11}$ interference) 
\cite{Anisovich_15} (black, solid).}
\end{figure}

\begin{figure*}[t]
\centerline{
\resizebox{\textwidth}{!}{%
    \includegraphics[height=0.04\textheight]{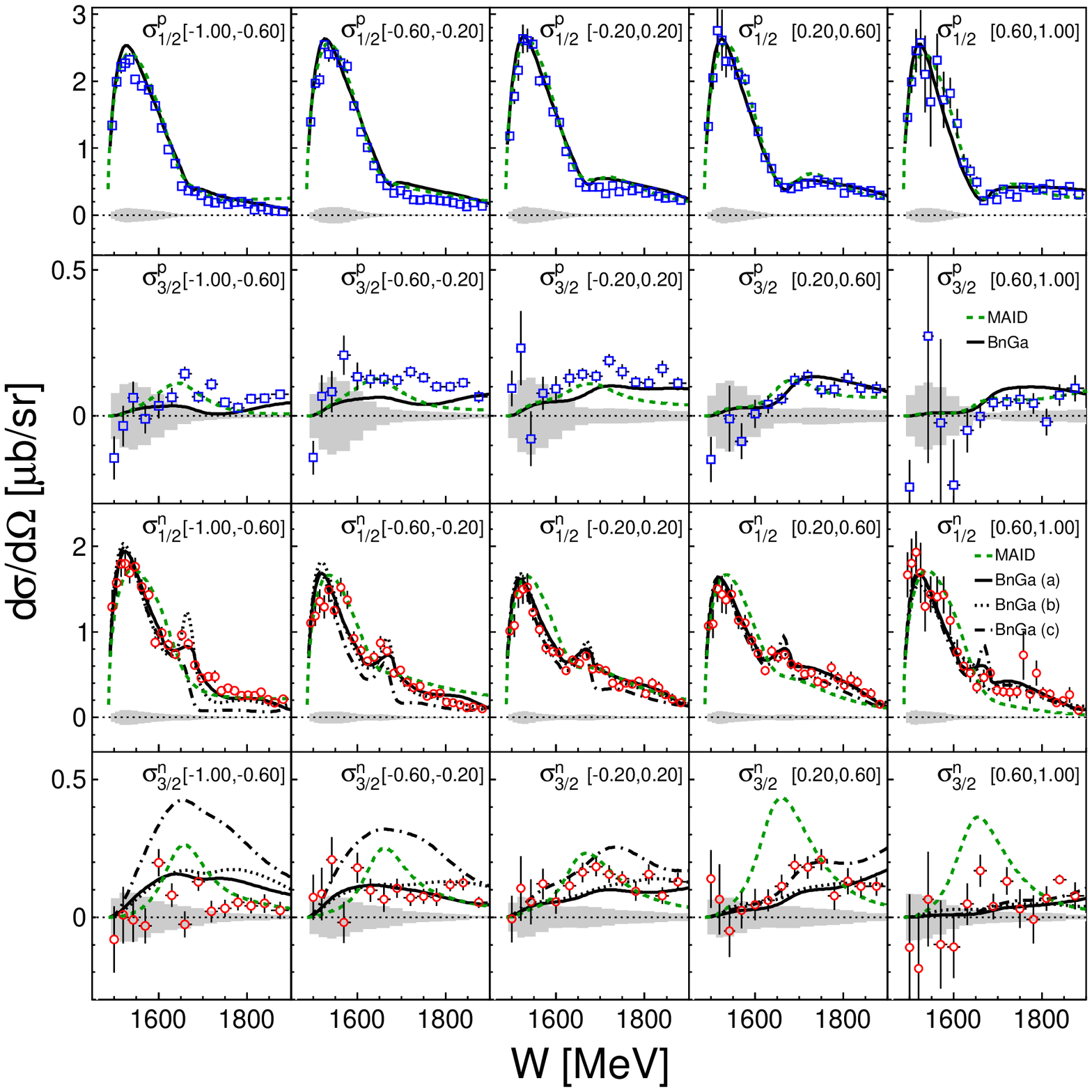}
    \includegraphics[height=0.04\textheight]{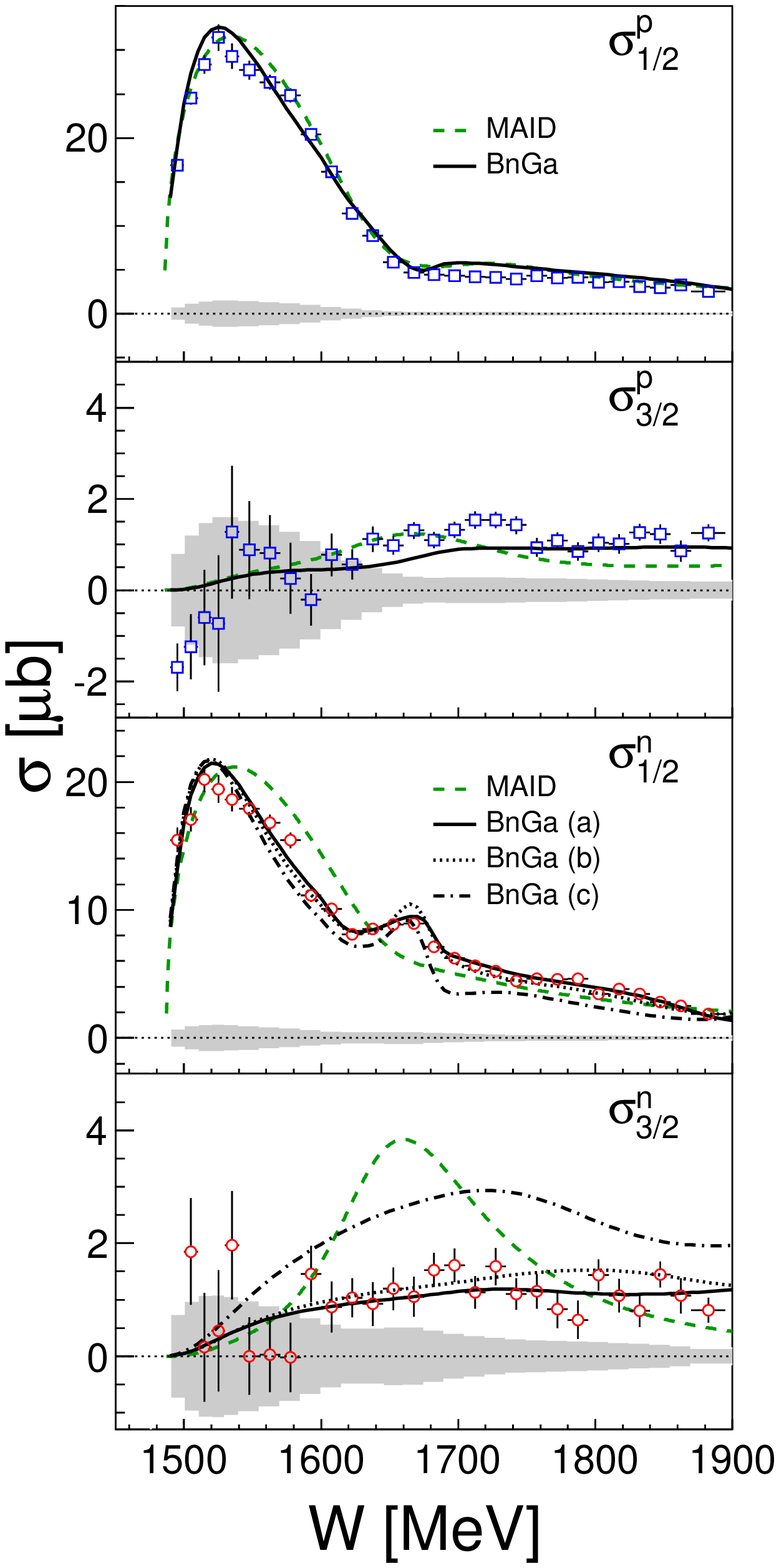}
    }}
\caption{\label{fig:fig1} (Color online) Excitation functions for $\sigma_{1/2}$ and $\sigma_{3/2}$ for five 
cos($\theta_{\eta}^{\star}$) bins (cos($\theta$) ranges given in figure). 
Top two rows: results for $\gamma p\rightarrow p\eta$ (blue squares).
Bottom two rows: $\gamma n\rightarrow n\eta$ (red circles). 
Gray shaded areas: systematic uncertainties.
Curves: model predictions from MAID (dashed green) \cite{Chiang_02}, 
BnGa(a) (model version with interference in $S_{11}$ wave, solid black) \cite{Anisovich_15}, 
BnGa(b) (model with narrow $P_{11}$ resonance with positive $A_{1/2}$ coupling, dotted black) \cite{Anisovich_15}, 
and BnGa(c) (narrow $P_{11}$ resonance with negative $A_{1/2}$ coupling, dash-dotted, black) \cite{Anisovich_15}.
Right hand side: total cross sections.
}
\end{figure*}

The only complication resulted from the contribution from nucleons bound in the unpolarized carbon (and oxygen) 
nuclei in the butanol target. This background contributes only in the denominator of Eq.~(\ref{eq:E}).   
It was determined from a measurement with a carbon foam target (which had identical geometry and density to 
the butanol target) and subtracted. Both measurements (butanol and carbon target) were normalized absolutely 
to photon fluxes, target surface densities, and detection efficiencies. 

The double polarization observable $E$ for $\eta$ mesons in coincidence with recoil protons and neutrons 
is shown in Fig.~\ref{fig:E}. The systematic uncertainty was estimated from the uncertainty of the target 
($\pm10\%$) and photon beam polarization ($\pm2.7\%$). In addition, there is a small uncertainty related 
to the subtraction of the carbon background (all other uncertainties e.g. from detection efficiencies cancel 
to a large extent in the ratio of Eq.~\ref{eq:E}). This uncertainty was estimated from the precision 
of the photon flux measurements and the determination of the target surface densities.
It is on the order of 2.5\% and was added quadratically to the polarization degree uncertainties.
As a cross check for the correct subtraction of the carbon background an analysis was done for which
the denominator of the ratio in Eq.~\ref{eq:E} was replaced by $2\sigma_0$, where $\sigma_0$ is the 
unpolarized total cross section measured with a liquid deuterium target (so that no subtraction of carbon 
data is necessary). The data for $\sigma_0$ were taken from \cite{Werthmueller_14}. The average deviation 
between the analyses using the carbon subtracted butanol or the liquid deuterium data in the denominator
was 2.25\% for recoil neutrons and 2.1\% for recoil protons. For the latter, only data above $W$=1.6~GeV 
were used for the comparison because for lower energies the detection efficiency for recoil protons 
(which cancels as long as Eq.~\ref{eq:E} is used with the carbon subtracted butanol data) could not be 
determined precisely enough for a comparison to the results of \cite{Werthmueller_14} on an absolute scale.  

The neutron data are in quite good agreement with the results from the BnGa model \cite{Anisovich_15}
and clearly rule out the MAID predictions \cite{Chiang_02}. The disagreement between measurement and MAID
prediction can be easily traced to an unrealistically large contribution of the $N(1675)5/2^-$ state in the 
MAID model. 

\begin{figure}[t]
\centerline{\resizebox{0.38\textwidth}{!}{\includegraphics{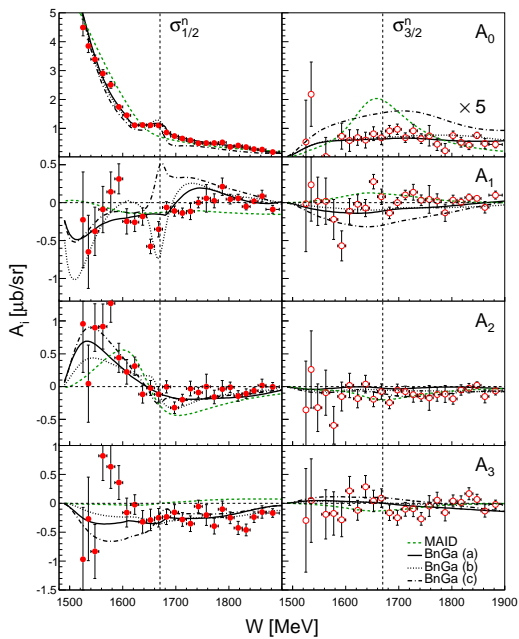}}}
\caption{\label{fig:Leg} (Color online) Legendre coefficients of the angular distributions of $\sigma_{1/2}$ 
and $\sigma_{3/2}$ for the reaction $\gamma n\rightarrow n\eta$. Experimental 
results (red circles) and model predictions by MAID \cite{Chiang_02} and BnGa \cite{Anisovich_15}.
Same notation as in Fig.~\ref{fig:fig1}. The vertical dashed lines at $W=1685$~MeV indicate the
position of the narrow structure. The results for $A_0$, $\sigma_{3/2}$ are up-scaled by factor of 5. 
}
\end{figure}

The helicity dependent cross sections $\sigma_{1/2}$ and $\sigma_{3/2}$ can be extracted as
\begin{equation}
\sigma_{1/2} = \sigma_0(1+E),~~~\sigma_{3/2} = \sigma_0(1-E),
\label{eq:heli}
\end{equation}
from the asymmetry $E$ and the unpolarized cross section $\sigma_0$. For the latter the results from
\cite{Werthmueller_14} were used. The results are summarized in Figs.~\ref{fig:fig1} and 
\ref{fig:Leg}. The systematic uncertainties for $E$ were propagated into Eq.~\ref{eq:heli}.
The overall systematic uncertainty for the scale of $\sigma_0$ from Ref.~\cite{Werthmueller_14}
is on the order of 7 - 15\%. It is also possible to construct $\sigma_{1/2}$ and $\sigma_{3/2}$
directly from the data measured with the butanol target after subtraction of the carbon background
without using input from the independent measurement of the unpolarized cross section. For the measurement
with recoil neutrons excellent agreement was found for all energies and cm angles of the $\eta$,
for recoil protons deviations occurred for $W< 1.6$~GeV due to the known inaccuracies of the proton
detection efficiency. 

Fig.~\ref{fig:fig1} shows the excitation functions for five bins of cos($\theta_{\eta}^{\star}$) 
($\theta_{\eta}^{\star}$ polar angle in the photon-nucleon center-of-momentum (cm) frame) and the 
total cross sections in comparison to the predictions from the MAID \cite{Chiang_02} and BnGa \cite{Anisovich_15}
models. For protons and neutrons, contributions from the helicity-3/2 amplitude are small, which means that 
nucleon resonances with $J\geq 3/2$ contribute little. For the proton target, the $\sigma_{1/2}$ results are 
in good agreement with model predictions. The small $\sigma_{3/2}$ part is in reasonable agreement with model
results. Details like the contribution of the $N(1720)3/2^+$ state (a small enhancement with respect to the model
results may be visible in the total $\sigma_{3/2}$ cross section in this energy range) will be subject to more
refined partial wave analysis.   

The results for the quasi-free neutron establish that the narrow structure around 
$W\approx 1.67$~GeV, listed as tentative N(1685) state in RPP, appears only in the helicity-1/2 part 
of the reaction. This means that it is almost certainly related to $J=1/2$ contributions ($S_{11}$ 
and/or $P_{11}$ partial waves). Although excited nucleon states with $J\geq 3/2$ can also contribute 
to helicity-1/2, it is unlikely that they contribute {\it only} to helicity-1/2. 
The RPP \cite{PDG_14} lists only one state up to excitation energies of 2~GeV for which 
the helicity coupling $A_{1/2}$ is larger than $A_{3/2}$ (the $N(1875)3/2^+$ for the proton, 
but even in that case within uncertainties $A_{3/2}$ could be larger). There is no example
for such a state for which the helicity-3/2 contribution is negligible compared to
helicity-1/2. Since no trace of the structure is observed in helicity-3/2, a contribution 
from $J\geq 3/2$ states is highly unlikely.

As mentioned above, a large contribution of the $N(1675)5/2^-$ state, as in the
MAID model, was ruled out. In addition, the BnGa model with a narrow $P_{11}$ resonance with negative
coupling disagrees with the experimental results, while the other two BnGa model versions
give similar results.
The angular distributions have been fitted with third order Legendre expansion
to allow for a more detailed comparison to model predictions:
\begin{equation}
\frac{d\sigma}{d\Omega}(W,\mbox{cos}(\theta_{\eta}^{\star})) = \frac{q_{\eta}^{\star}(W)}{k_{\gamma}^{\star}(W)}
\sum_{i=0}^{3} A_i(W) P_i(\mbox{cos}(\theta_{\eta}^{\star}))\,,
\end{equation}
where $q_{\eta}^{\star}$ and $k_{\gamma}^{\star}$ are the $\eta$ and photon momenta in the cm frame, 
respectively. The results are shown in Fig.~\ref{fig:Leg}. The $A_1$ coefficient for the $\sigma_{1/2}$ 
cross section is very interesting. An interference between a $P_{11}$ wave and the dominant $S_{11}$ 
wave results in a cos($\theta_{\eta}^{\star}$) term in the angular distribution,
which is reflected in the $A_{1}$ coefficient. Depending on the sign of the interference term,
a narrow $P_{11}$ resonance will result in a sharp positive or negative peak in $A_{1}$, as shown
by the model curves in Fig.~\ref{fig:Leg}, while interference effects in the $S_{11}$ wave produce 
different patterns. The results clearly rule out the model version with a negative $P_{11}$-$S_{11}$ 
interference sign. However, the model results with a positive interference sign of $P_{11}$ 
and $S_{11}$ are more similar to the measured data than the predictions without the addition of 
a narrow $P_{11}$ state. 

In summary, the double polarization observable $E$ and the related helicity dependent cross sections
$\sigma_{1/2}$ and $\sigma_{3/2}$ were measured for the first time for photoproduction of $\eta$ 
mesons on quasi-free nucleons using a circularly polarized photon beam and a longitudinally
polarized target. The measurement provided data of excellent quality, which are important input 
for future partial wave analysis of photoproduction of $\eta$ mesons off nucleons. 
Here, we report one striking finding about the nature of the narrow structure 
previously observed in the $\gamma n\rightarrow n\eta$ reaction. The results have unambiguously
established that this structure is related to the helicity-1/2 amplitude and a comparison of the
angular dependence to different model predictions favors a scenario with a contribution from a narrow 
$P_{11}$ resonance.

\begin{acknowledgments}
We wish to acknowledge the outstanding support of the accelerator group and operators of MAMI. 
This work was supported by Schweizerischer Nationalfonds (200020-156983, 132799, 121781, 117601), 
Deutsche For\-schungs\-ge\-mein\-schaft (SFB 443, SFB 1044, SFB/TR16), the INFN-Italy, 
the European Community-Research Infrastructure Activity under FP7 programme (Hadron Physics, 
grant agreement No. 227431), 
the UK Science and Technology Facilities Council (ST/J000175/1, ST/G008604/1, ST/G008582/1,ST/J00006X/1, and 
ST/L00478X/1), 
the Natural Sciences and Engineering Research Council (NSERC, FRN: SAPPJ-2015-00023), Canada. This material 
is based upon work also supported by the U.S. Department of Energy, Office of Science, Office of Nuclear 
Physics Research Division, under Award Numbers DE-FG02-99-ER41110, DE-FG02-88ER40415, and DE-FG02-01-ER41194 
and by the National Science Foundation, under Grant Nos. PHY-1039130 and IIA-1358175.
\end{acknowledgments}

\end{document}